\documentclass[preprint,aps,nofootinbib,tightenlines,amsfonts]{revtex4}
\usepackage{epsfig}
\usepackage{bm}

\def\OMIT#1{{}}

\def\epp{\epsilon^\prime}
\def\ep{\epsilon}

\begin{document}

\begin{figure}[!t]
\vskip -1.5cm
\leftline{{\epsfxsize=1.5in \epsfbox{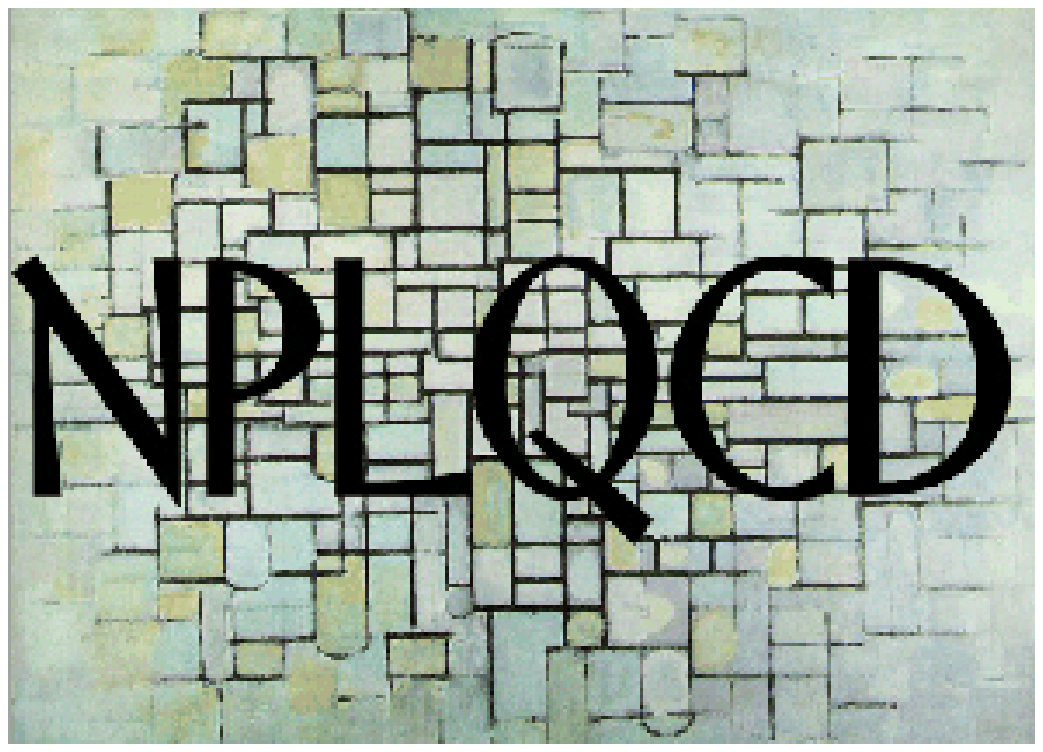}}}
\end{figure}

\preprint{\vbox{ \hbox{NT@UW-04-017} }}

\vphantom{}

\title{Nucleon Properties at Finite Volume: the $\epp$-Regime } 
\author{William Detmold and Martin J.  Savage}
\affiliation{ Department of Physics, University of Washington,
  Seattle, WA 98195-1560, U.S.A.}

\vphantom{} \vskip 0.5cm
\begin{abstract} 
  \vskip 0.5cm
\noindent 
We study the properties of the nucleon in highly asymmetric volumes
where the spatial dimensions are small but the time dimension is large
in comparison to the inverse pion mass. To facilitate power-counting
at the level of Feynman diagrams, we introduce $\epp$-power-counting
which is a special case of Leutwyler's $\delta$-power-counting.  Pion
zero-modes enter the $\epp$-counting perturbatively, in contrast to
both the $\ep$- and $\delta$-power-countings, since $m_q \langle
q\overline{q}\rangle V$ remains large. However, these modes are
enhanced over those with non-zero momenta and enter at lower orders in
the $\epp$-expansion than they would in large volume chiral
perturbation theory.  We discuss an application of $\epp$-counting by
determining the nucleon mass, magnetic moment and axial matrix element
at the first nontrivial order in the $\epp$-expansion.  \vskip 1.0cm
\leftline{\today}
\end{abstract}

\maketitle

\vfill\eject

\section{Introduction}

Lattice QCD is the only known way of computing strong interaction
observables rigorously from QCD.  Even if (unquenched) lattice QCD
simulations were to be performed at the physical values of the
light-quark masses, $m_q$, extrapolations would be required to obtain
information about nature.  As simulations are performed in lattice
volumes of finite size, $V=L_4 \times L^3$ (where $L_4$ is the length
of the time-dimension and $L$ is the length of each spatial-dimension
which are taken to be the same for simplicity), an extrapolation to
the infinite-volume limit, $L,\, L_4 \to\infty$, is required. In
addition, as the lattice-spacing ($a$) is necessarily finite, an
extrapolation to vanishing lattice spacing, $a\to0$, is essential.
Currently, an extrapolation in $m_q$ from those used in the
simulations down to those of nature is also required.

In order to perform these extrapolations, it is necessary to construct
the effective field theory (EFT) appropriate for the lattice
simulations.  The dependence upon lattice spacing, lattice volume and
quark masses computed in the EFT can be used to extract the
coefficients of local operators that contribute to the observable of
interest (at any given order in the EFT expansion). One can then use
the EFT to compute the observable at the physical values of the quark
masses, in the $m_\pi L,\,m_\pi L_4\gg 1$ limit and in the
$a\Lambda_\chi \ll 1$ (where $\Lambda_\chi$is the scale of chiral
symmetry breaking) limit.  Further, once appropriate counterterms have
been determined, observables that are significantly more difficult to
compute directly from lattice simulations can be computed with the
EFT.

For simple one-body observables (we do not discuss the two-particle
scattering amplitude or higher-body interactions on the lattice in
this work) in large volumes, deviations from the infinite-volume limit
are exponentially suppressed by factors of $m_\pi L,\, m_\pi L_4\gg 1$
(the scale is set by $m_\pi$ since the pions are the lightest hadrons)
and can be computed with the $p$-power-counting of infinite-volume
chiral perturbation theory~\cite{Luscher:1985dn}.  However, in small
volumes with $m_\pi L,\, m_\pi L_4\sim\ep\ll 1$ but with $m_q \langle
q\overline{q}\rangle L^3 L_4\sim 1$ (where $\langle
q\overline{q}\rangle$ is the infinite-volume quark condensate) one is
in the $\ep$-regime, where the small expansion parameter is $\ep$ and
not $p/\Lambda_\chi$. Here, the contribution from the pion zero-modes
are non-perturbative and must be resummed to all
orders~\cite{Gasser:1987ah,Hansen:1990un,Hansen:1990yg,Hansen:1990kv,Hasenfratz:1989pk,Hasenfratz:1989ux,Leutwyler:1992yt}.
This regime has been explored both numerically and analytically in the
light meson sector (e.g.
Refs.~\cite{Giusti:2004yp,Giusti:2003gf,Giusti:2002sm,Bietenholz:2003bj,Durr:2000ei,Smilga:2000ek}),
and efforts in the single nucleon~\cite{paco}, and
heavy-meson~\cite{dlin} sectors are underway.

In this work we point out that for heavy-hadrons, such as nucleons, it
is useful to explore the behavior of observables in highly asymmetric
volumes where $m_\pi L\ll1$ and $m_\pi L_4\gg1$. The underlying
reason for this is simply that such objects are near their mass-shell
when their kinetic energy and three-momentum are related via $E \sim
|{\bf k}|^2$. Such asymmetric volumes were first analyzed by Leutwyler
\cite{Leutwyler:1987ak} in the context of pion dynamics. Leutwyler
introduced $\delta$-power-counting for which $m_\pi L_4 \sim \delta^0$
while $m_\pi L\sim \delta^2$ and $\Lambda_\chi L \sim 1/\delta$. This
is similar to the $\ep$-regime as $m_q \langle q\overline{q}\rangle
L^3 L_4\sim \delta^0$ and the pion zero modes must be summed
non-perturbatively.  If $L_4$ becomes large, there is an additional
small parameter, $1/L_4$, and it is convenient to define a further
power-counting describing this regime.  We define a new expansion
parameter $\epp$, and assign\footnote{There is some freedom in the
  counting of $m_\pi$ and $L_4$; we simply require that $m_q \langle
  q\overline{q}\rangle L^3 L_4\gg 1$.}
\begin{eqnarray}
\Lambda_\chi L & \sim & {1 \over \epp}
\ \ ,\ \ 
\frac{m_\pi^2}{ \Lambda_\chi^2} \sim\ \frac{m_q}{\Lambda_\chi }\ \sim\ {\epp}^4
\ \ ,\ \ 
\Lambda_\chi L_4\ \sim\ {1\over {\epp}^\alpha}
\ \ ,\ \ 
\label{eq:eppscaling}
\end{eqnarray}
in the power-counting, which provides $m_\pi L\ll1$, $m_\pi L_4\gg 1$
and $\Lambda_\chi L, \,\Lambda_\chi L_4\gg 1$ for $\alpha\gg 2$.  In
contrast to their behavior in the $\ep$-regime, in the $\epp$-regime,
the pion zero-modes (modes with zero spatial momentum) are
perturbative as $m_q \langle q\overline{q}\rangle L^3 L_4\sim
{\epp}^{(1-\alpha)}$ is large. As for the $m_\pi L,\, m_\pi L_4 \gg 1$
regime, the vacuum structure in the $\epp$-regime is perturbatively
close to that at infinite volume. Correspondingly, the $\epp$-regime
is that part of the $\delta$-regime where $m_q \langle
q\overline{q}\rangle L^3 L_4$ is large.

Here we shall take all three spatial dimensions to be of order $\epp$
but the same modified power-counting will emerge even if only one of
spatial directions is small.  In what follows we will work in the
$\alpha=\infty$ limit in which the time-direction is infinite and will
assume the lattice spacing vanishes. In principle deviations from
these limits can be included perturbatively.

Finite volume effects on the properties of heavy hadrons computed in
lattice QCD have recently been investigated for $L_4=\infty$ using the
standard power-counting of chiral perturbation theory valid for $m_\pi
L\gg 1$ (see Refs.~\cite{Beane:2004tw, Beane:2004rf, Arndt:2004bg, Leinweber:2001ac,
  Young:2004tb, Detmold:2002nf}). The $\epp$-counting introduced here
allows for consistent calculations with $m_\pi L\ll1$, and in what
follows we calculate the nucleon mass, magnetic moment and axial
matrix element in this regime.

\section{Nucleons at Finite Volume}

At leading order in the $p$-expansion, the Lagrange density describing
the low-energy dynamics of the nucleons, $\Delta$'s and pions
(pseudo-Goldstone bosons) that is consistent with the spontaneously
broken $SU(2)_L\otimes SU(2)_R$ approximate chiral symmetry of QCD
is~\cite{Jenkins:1990jv,Jenkins:1991ne}
\begin{eqnarray}
{\cal L} & = & 
\overline{N}\ iv\cdot {\cal D} \ N\ - \ 
\overline{T}_\mu\ iv\cdot {\cal D}\  T^\mu
\ +\ 
\Delta\ \overline{T}_\mu\ T^\mu
\nonumber\\
& + & 
{f^2\over 8} {\rm Tr}\left[ \partial_\mu\Sigma^\dagger \ \partial^\mu\Sigma\
\right]
\ +\ \lambda\ {f^2\over 4}  {\rm Tr}\left[ m_q \Sigma^\dagger\ +\ {\rm h.c.}\
\right]
\nonumber\\
& + &  
2 g_A\  \overline{N} S^\mu  A_\mu N
\ +\  
g_{\Delta N}\ 
\left[\ 
\overline{T}^{abc,\nu}\  A^d_{a,\nu}\  N_b \ \epsilon_{cd} 
\ +\ {\rm h.c.}
\ \right]
\ +\ 
2 g_{\Delta\Delta}\  \overline{T}_\nu S^\mu  A_\mu T^\nu
\ \ \ ,
\label{eq:lagstrong}
\end{eqnarray}
where ${\cal D}$ is the chiral covariant derivative, $N$ is the
nucleon field operator, and $T^\mu$ is the Rarita-Schwinger field
containing the quartet of spin-${3\over 2}$ $\Delta$-resonances as
defined in heavy-baryon $\chi$PT~\cite{Jenkins:1990jv,
  Jenkins:1991ne}.  The mass difference between the
$\Delta$-resonances and the nucleon is $\Delta$ (taken to be $\sim
m_\pi$), and $g_A\sim1.26$, $g_{\Delta N}$ and $g_{\Delta\Delta}$ are
the (infinite volume, chiral limit) axial couplings between the
baryons and pions.\footnote{In the naive constituent quark model,
  $|g_{\Delta N}|/g_{A}=6/5$ and $|g_{\Delta\Delta}|/g_A=9/5$.}
$S_\mu$ is the covariant spin vector \cite{Jenkins:1990jv,
  Jenkins:1991ne}, and $v_\mu$ is the heavy-baryon four-velocity, with
$v^2=1$.  Pions appear in eq.~(\ref{eq:lagstrong}) through $\Sigma$
and $A_\mu$ which are defined to be
\begin{eqnarray}
\Sigma & = & \exp\left({2 i M \over f}\right)
\ =\ \xi^2
,\ 
A^\mu \ =\  {i\over 2}\left(\ \xi\partial^\mu\xi^\dagger
-
\xi^\dagger\partial^\mu\xi \ \right)
,\ 
M \ = \ \left(\matrix{\pi^0/\sqrt{2} & \pi^+\cr \pi^- & -\pi^0/\sqrt{2}}\right)
\ ,\,
\label{eq:mesons}
\end{eqnarray}
and $f\sim132$~MeV is the pion decay constant.  With this leading
order Lagrange density one can determine the leading chiral loop
corrections to observables in the single nucleon sector.

\subsection{The Nucleon Mass}

\subsubsection{Nucleon Mass in the $p$-expansion}

\begin{figure}[!t]
\centerline{{\epsfxsize=4.0in \epsfbox{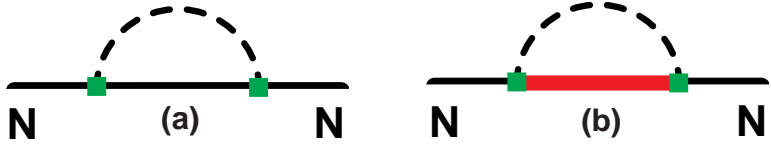}}} 
\vskip 0.15in
\noindent
\caption{\it 
  One-loop graphs that give the leading contributions to the nucleon
  mass in the large-volume limit.  The solid, thick-solid and dashed
  lines denote a nucleon, $\Delta$-resonance, and a pion,
  respectively.  The solid-squares denote an axial coupling given in
  eq.(\ref{eq:lagstrong}).}
\label{fig:masses}
\vskip .2in
\end{figure}

The nucleon mass calculated at the first nontrivial order in the
$p$-expansion in the isospin limit is known to
be~\cite{Jenkins:1990jv,Jenkins:1991ne,Bernard:1993nj}
\begin{eqnarray}
M_N(\infty) & = & M_0 - 2 \overline{m}\left( \alpha_M+2\sigma_M\right)
- {1\over 8\pi f^2}\left[\ {3\over 2} g_A^2 m_\pi^3
\ +\ {4 g_{\Delta N}^2\over 3\pi} F_\pi\ \right]
\ \ \ ,
\label{eq:Nmass}
\end{eqnarray}
where $M_0$ is the chiral limit nucleon mass and the constants
$\alpha_M$ and $\sigma_M$ are coefficients in the Lagrange density
\begin{eqnarray}
{\cal L}_M & = & 2\alpha_M \overline{N} {\cal M}_+ N\ +\ 
2\sigma_M  \overline{N}  N {\rm Tr}\left[ {\cal M}_+ \right]
\ \ \ ,
\label{eq:lagmass}
\end{eqnarray}
where ${\cal M}_+={1\over 2}\left( \xi^\dagger m_q\xi^\dagger + \xi
  m_q\xi \right)$.  The function $F_\pi$ is given by
\begin{eqnarray}
F_\pi & = & 
\left(m_\pi^2-\Delta^2\right)\left(
\sqrt{\Delta^2-m_\pi^2} \log\left({\Delta -\sqrt{\Delta^2-m_\pi^2+i\epsilon}\over
\Delta +\sqrt{\Delta^2-m_\pi^2+i\epsilon}}\right)
-\Delta \log\left({m_\pi^2\over\mu^2}\right)\ \right)
\nonumber\\
&& 
-{1\over 2}\Delta m_\pi^2 \log\left({m_\pi^2\over\mu^2}\right)
\ \ \ . 
\label{eq:massfun}
\end{eqnarray}
The three terms in eq.~(\ref{eq:Nmass}) scale as $p^0$, $p^2$ and
$p^3$ in the $p$-expansion, respectively, and the loop contribution
entering at order $p^3$ results from the diagrams shown in
Fig.~\ref{fig:masses}.

The modifications to the nucleon mass in a volume with all dimensions
large compared to the inverse pion mass have been recently computed by
Beane~\cite{Beane:2004tw}.  It is found that the leading finite-volume
corrections in the $m_\pi L,\, m_\pi L_4 \gg 1$ limit also arise from
the one-loop diagrams that give the order $p^3$ contributions to the
nucleon mass in the infinite-volume limit~\cite{Beane:2004tw},
\begin{eqnarray}
\label{eq:MN}
M_N (L) & = & M_N (\infty)\ -\ {1\over 12\pi^2}
\ \left[\ {9\over 2} {g_A^2\over f^2} {\cal K}(0)
\ +\ 4 {g_{\Delta N}^2\over f^2} {\cal K}(\Delta)\ \right]
\ \ \ ,
\end{eqnarray}
where the function ${\cal K}(\Delta)$ is
\begin{eqnarray}
{\cal K}(\Delta) & = & 
\sum_{{\bf n}\ne {\bf 0}}\ 
\int_{m_\pi}^\infty\ d\beta
{\beta^3\over\sqrt{\beta^2+\Delta^2-m_\pi^2}}
\left[\ {1\over \beta L |{\bf n}|} K_1 (\beta L |{\bf n}|) - 
K_0 (\beta L |{\bf n}|)\ \right]
\ \ \ ,
\label{eq:kdelta}
\end{eqnarray}
where the $K_\nu (z)$ are modified Bessel functions of order $\nu$.
For $\Delta=0$ this expression reduces to~\cite{Beane:2004tw}
\begin{eqnarray}
{\cal K}(0) & = & 
-{\pi m_\pi^3\over 2}
\sum_{{\bf n}\ne {\bf 0}}\ 
{1\over m_\pi L |{\bf n}|} e^{-m_\pi L |{\bf n}|}
\ \ \ .
\end{eqnarray}

\subsubsection{Nucleon Mass in the $\epp$-expansion}

The order at which operators contribute in the $\epp$-expansion is
different from the order at which they contribute in the
$p$-expansion.  The local operators given in eq.~(\ref{eq:lagmass})
contribute at order $p^2$ in the $p$-expansion, but contribute at
order ${\epp}^4$ in the $\epp$-expansion.  Power-counting the loop
contributions is a little more complicated than in the $p$-regime, and
is similar in some respects to the counting in the $\ep$-regime.
Since $|{\bf p}|\sim1/L\sim\epp$ and $m_\pi\sim{\epp}^2$, the spatial
momentum zero-mode and non-zero modes must be treated separately as
their counting is different,
\begin{eqnarray}
  \label{eq:1}
  \frac{i}{k_0^2-|{\bf k}|^2-m_\pi^2 +i \eta}\sim  \left\{\begin{array}{ll}
    L^2\sim {\epp}^{-2} \quad&{\text{ non-zero mode}}\\ 
    m_\pi^{-2}\sim {\epp}^{-4} \quad&{\text{ zero mode }} (|{\bf k}|=0)
\end{array}\right.
\quad .
\end{eqnarray}
In calculating loop diagrams one picks out the pole at
$k_0=\sqrt{|{\bf k}|^2+m_\pi^2}+i\eta$, so $k_0\sim|{\bf k}|\sim\epp$
if $|{\bf k}|\ne0$ and $k_0\sim m_\pi \sim{\epp}^2$ otherwise.

For the one-loop diagrams contributing to the nucleon mass, the
momentum zero-mode does not contribute due to the derivative couplings
in eq.~(\ref{eq:lagstrong}) and therefore diagram (a) in
Fig.~\ref{fig:masses} becomes
\begin{eqnarray}
{\rm loop} &\sim &
{1\over L^3}\ \sum_{{\bf k}}\ \int {dk_0\over 2\pi}\ 
{1\over k_0^2 - |{\bf k}|^2}\ {1\over k_0} \ |{\bf k}|^2
\sim\  {\epp}^3\ \epp \ {\epp}^{-2}\  {\epp}^{-1}\  {\epp}^2
\ \sim \ {\epp}^3
\ \ .
\end{eqnarray}
Since the $k_0$ integral gives $k_0\sim |{\bf k}| \sim \epp$, and
$\Delta\sim m_\pi \sim {\epp}^2$, 
\begin{eqnarray}
  \label{eq:2}
  \frac{1}{k_0-\Delta}=\frac{1}{k_0}+\frac{\Delta}{k_0^2}+{\cal O}(\epp)\,,
\end{eqnarray}
and the leading contribution from loops with $\Delta$-intermediate
states can be found by setting $\Delta=0$.  After some elementary
manipulations, it is easy to show that in the $\epp$-regime, the
function ${\cal K}(\Delta)$ in eq.~(\ref{eq:kdelta})
becomes~\footnote{ In particular we have used the relation
\begin{eqnarray*}
\sum_{{\bf n}\ne {\bf 0}}\ 
{1\over |{\bf n}|} e^{-z |{\bf n}|}
& = & 
\sum_{{\bf n}\ne {\bf 0}}\ 
{1\over |{\bf n}|} e^{- |{\bf n}|}
+{4\pi\over z^2}-4\pi-1
-
\sum_{{\bf p}\ne {\bf 0}}\ 
{4\pi (z^2-1)\over (z^2 + 4\pi^2 |{\bf p}|^2)  (1 + 4\pi^2 |{\bf p}|^2)}
+z
\ \ \ ,
\end{eqnarray*}
and define divergent sums through dimension regularization, e.g.
\begin{eqnarray*}
  \sum_{{\bf n}\ne {\bf 0}}\ 
{1\over |{\bf n}|}= \lim_{z\to0}\left[\sum_{{\bf n}\ne {\bf 0}}\ 
{1\over |{\bf n}|} e^{-z |{\bf n}|}
-\frac{4\pi}{z^2}\right]\,.
\end{eqnarray*}
}
\begin{eqnarray}
{\cal K}(\Delta)
& = & 
-{2\pi^2\over L^3}\left[\ 
1\  +\  
{\Delta L\over 2\pi} c_1
\ +\ 
{ (m^2-\Delta^2)L^2\over 4\pi^2}
c_1
 \right]
\ +\ 
{\cal O}({\epp}^6)
\ \ \ ,
\end{eqnarray}
where $c_1=-2.8372974$ is the same geometric number that appears in
the $1/L$ expansion of the lowest two-particle continuum energy-level
in a finite volume~\cite{Luscher:1986pf}.

{}From this expression, we see that the one-loop diagrams that produce
the order $p^3$ contribution in eq.~(\ref{eq:Nmass}) contribute at
order ${\epp}^3$ in the $\epp$-expansion, and provides the leading
volume dependence in this regime.  Combining this result with
eq.~(\ref{eq:MN}), we arrive at
\begin{eqnarray}
M_N (L) & = & M_0
\ +\ {1\over f^2 L^3}\ \left[\ {3\over 4} g_A^2 + {2\over 3} g_{\Delta N}^2
\ \right]
\ +\ {\cal O}({\epp}^4)
\ \ \ .
\end{eqnarray}

A detailed study of the volume dependence of the nucleon mass in the
$\epp$-regime (including higher order corrections) would provide a
clean method of determining this particular combination of axial
couplings. This may provide the simplest way of determining the
transition axial coupling, $g_{N\Delta}$, as it cannot be reliably
determined from analysis of a three-point function at the physical
values of the quark masses.

\subsection{The Nucleon Magnetic Moment}

\subsubsection{Nucleon Magnetic Moment  in the $p$-expansion}

\begin{figure}[!t]
\centerline{{\epsfxsize=4.0in \epsfbox{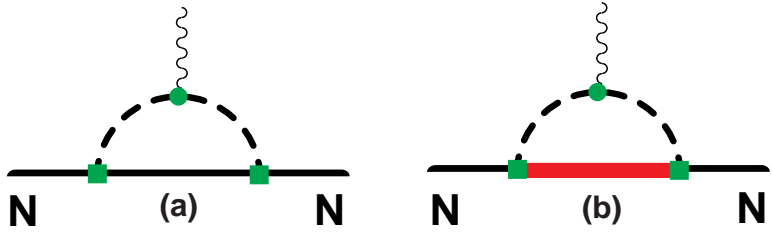}}} 
\vskip 0.15in
\noindent
\caption{\it 
  One-loop graphs that contribute to the nucleon magnetic moment.  The
  solid, thick-solid and dashed lines denote a nucleon,
  $\Delta$-resonance, and a meson, respectively.  The solid-squares
  denote an axial coupling given in eq.(\ref{eq:lagstrong}) and the
  solid-circles denote a leading-order electromagnetic interaction.}
\label{fig:magmoms}
\vskip .2in
\end{figure}

The magnetic moment of the nucleon has been computed in two-flavor and
three-flavor infinite-volume chiral perturbation
theory~\cite{Caldi:ta,Jenkins:1992pi,Meissner:1997hn,Durand:1997ya},
and in the isospin-limit of the two-flavor case is known to be
\begin{eqnarray}
\hat\mu(\infty) & = & 
\mu_0\ +\ \mu_1 \tau^3 
\ -\ {M_N\over 4\pi f^2}\left[\ g_A^2 \ m_{\pi} 
+ {2\over 9}\ g_{\Delta N}^2\  {\cal F}_{\pi}\ \right] \tau^3 
\ \ \ ,
\label{eq:magmomsQCD}
\end{eqnarray}
where 
\begin{eqnarray}
\pi {\cal F}_\pi
& = & \sqrt{\Delta^2-m_\pi^2}\log\left({\Delta-\sqrt{\Delta^2-m_\pi^2+i\epsilon}
\over \Delta+\sqrt{\Delta^2-m_\pi^2+i\epsilon}}\right)
\ -\ \Delta\log\left({m_\pi^2\over\mu^2}\right)
\ \ \ .
\label{eq:magfun}
\end{eqnarray}
In the limit $\Delta\rightarrow 0$, ${\cal F}_\pi=m_\pi$.  The
quantities $\mu_{0,1}$ are the coefficients of the dimension-five
operators in the Lagrange density
\begin{eqnarray}
  \label{eq:mulag}
{\cal L}&=&\frac{e}{4M_N}F_{\mu\nu}\left(\mu_0
  \overline{N}\sigma^{\mu\nu}N +\mu_1
  \overline{N}\sigma^{\mu\nu}\tau_{\xi^+}^3 N\right) \,,
\end{eqnarray}
where $F_{\mu\nu}$ is the electromagnetic field-strength tensor and
$\tau_{\xi^+}^a= \frac{1}{2} \left(\xi^\dagger\tau^a\xi +
  \xi\tau^a\xi^\dagger\right)$.  The leading contribution to the
magnetic moment comes from these dimension-five operators and is of
order $p^0$, while the contributions from the one-loop diagrams shown
in Fig.~\ref{fig:magmoms} [the third term in
eq.~(\ref{eq:magmomsQCD})] are of order $p$.

Beane~\cite{Beane:2004tw} computed the leading finite-volume
corrections to the nucleon magnetic moment in the $p$-expansion for
$m_\pi L,\,m_\pi L_4\gg 1$, finding
\begin{eqnarray}
\hat\mu (L) & = & \hat\mu (\infty)\ +\ 
{M_N\over 6\pi^2 f^2}\ \left[\ g_A^2 {\cal Y}(0)\ +\ {2\over 9} g_{\Delta N}^2 
 {\cal Y}(\Delta)\ \right]\ \tau^3
\ \ \ ,
\end{eqnarray}
where the function $  {\cal Y}(\Delta)$ is given by
\footnote{The function ${\cal Y}(\Delta)$ is related to the function ${\cal
  K}(\Delta)$ defined in eq.~(\ref{eq:kdelta}) by
\begin{eqnarray*}
 {\cal Y}(\Delta) & = & -2\ {\partial  {\cal K}(\Delta)\over\partial m_\pi^2}
\ \ \ .
\end{eqnarray*}
}
\begin{eqnarray}
\label{eq:Ydelta}
 {\cal Y}(\Delta) & = & 
\sum_{{\bf n}\ne {\bf 0}}\ 
\int_{m_\pi}^\infty\ d\beta
{\beta\over\sqrt{\beta^2+\Delta^2-m_\pi^2}}
\left[\ 
3 K_0 (\beta L |{\bf n}|) - \beta L |{\bf n}| K_1 (\beta L |{\bf n}|)
\ \right]
\ \ \ ,
\end{eqnarray}
and in the $\Delta=0$ limit one has
\begin{eqnarray}
{\cal Y}(0) & = & 
-{\pi m_\pi\over 2}
\sum_{{\bf n}\ne {\bf 0}}\ 
\left[\ 1 - {2\over m_\pi L |{\bf n}|}\ \right] e^{-m_\pi L |{\bf n}|}
\ \ \ .
\end{eqnarray}

\subsubsection{Nucleon Magnetic Moment in the $\epp$-expansion}

The finite-volume contributions to the magnetic moment in the
$\epp$-regime behave quite differently from those of the nucleon mass.
The dimension-5 operators in eq.~(\ref{eq:mulag}) again give the
leading contribution, entering at order ${\epp}^0$.  Part of the
leading finite-volume correction, order $\epp$, comes from the
one-loop diagrams of Fig.~\ref{fig:magmoms} and it is straightforward
to calculate their contribution to the nucleon magnetic moment in the
$\epp$-regime.
\begin{figure}[!t]
\centerline{{\epsfxsize=1.75in \epsfbox{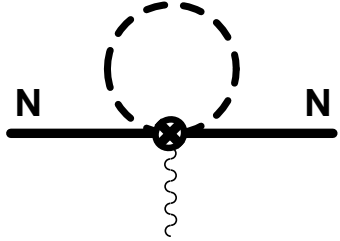}}}
\vskip 0.15in
\noindent
\caption{\it 
  One-loop operator insertion that contributes to the nucleon magnetic
  moment at next-to-leading order in the $\epp$-regime.  At large
  volume, this contribution is suppressed, contributing at order
  $p^2$.  The crossed-circle indicates the leading two-pion correction
  to the isovector electromagnetic current from eq.~(\ref{eq:mulag}).}
\label{fig:optad}
\vskip .2in
\end{figure}
However, at the same order in the $\epp$-expansion, the two-pion
contribution to the isovector magnetic moment operator generates the
tadpole loop diagram shown in Fig.~\ref{fig:optad}. This diagram
contributes at sub-leading order ($p^2$) in the $p$-expansion, but in
the $\epp$-regime, the modified counting of zero modes enhances this
to order $\epp$ if the pion has zero three-momentum. The infinite
volume integral corresponding to this diagram (defined in dimensional
regularization) is
\begin{eqnarray}
\label{eq:R1}
R_\pi & = &\mu^{4-n}\int {d^nq\over (2\pi)^n}\ 
{1\over q^2 - m^2_\pi + i \eta}\ \,.
\end{eqnarray}
In the $\epp$-regime, this becomes
\begin{eqnarray}
\label{eq:r1fv}
{1\over L^3} \sum_{{\bf q}}
\int dq_0\ {1\over q_0^2 - |{\bf q}|^2 -m_\pi^2 +i\eta}\ ,
\end{eqnarray}
and again, $q_0\sim|{\bf q}|\sim\epp$ generically, but for $|{\bf
  q}|=0$, $q_0\sim m_\pi\sim {\epp}^2$. Thus the contribution of this
diagram is order ${\epp}$ for zero-modes, and ${\epp}^2$ otherwise.

Taking these various contributions into account, we see that the
nucleon magnetic moment in the $\epp$-regime is
\begin{eqnarray}
\hat\mu (L) & = & \mu_0\ +\ \mu_1 \tau^3\left[1-\frac{1}{m_\pi f^2
    L^3}\right]  
\ +\ 
{M_N \ c_1\over 6\pi f^2 L}\ \left[\ g_A^2 \ +\ {2\over 9} g_{\Delta
    N}^2\ \right]\ \tau^3 
\ +\ {\cal O}({\epp}^2)
\ \ \ ,
\end{eqnarray}
and the leading volume dependence of the isovector magnetic moment is
$\sim 1/L,\, 1/(m_\pi L^3)$ occurring at order $\epp$. For the
isoscalar magnetic moment, the first correction occurs at order
${\epp}^2$.

The appearance of the mass of the pion in the denominator of one of
the leading corrections might be cause for worry.  However, one is not
able to take the $m_\pi\rightarrow 0$ limit of this result and remain
in the $\epp$-regime.  For small enough pion mass, $m_\pi L$ is no
longer of order $\epp$ and the perturbative series must be rearranged
to yield sensible results. One would move into the $\delta$-regime,
where the pion zero-modes would need to be resummed.

The above result has a number of interesting features. The first is
that pion mass dependent and pion mass independent volume corrections
enter at the same order. For a given volume, the numerical value of
$m_\pi L^2 \sim {\cal O}(1)$ can enhance or suppress different parts
of the correction.  A further, somewhat curious property of this
result is the strong dependence upon the geometry of the spatial
dimensions as is evidenced by the appearance of $c_1$ in the leading
correction.  In a study of the two-nucleon system in the presence of
background electroweak fields~\cite{Detmold:2004qn}, it was observed
that there are particular spatial geometries for which the coefficient
$c_1$ vanishes (e.g. in a volume with sides in the ratio
$1:1:3.72448$) and others where it is much larger than its symmetric
volume value (see also Ref.~\cite{Li:2003jn}). Thus, by changing the
shape of the spatial volume, the interplay of the different order
$\epp$ contributions can also be modified. In lattice calculations,
such freedom should enable one to eliminate or enhance the leading
corrections.

\subsection{The Axial Current Matrix Element in the Nucleon}

\subsubsection{ The Axial Current Matrix Element in the Nucleon  in
  the $p$-expansion}

\begin{figure}[!ht]
\centerline{{\epsfxsize=3.0in \epsfbox{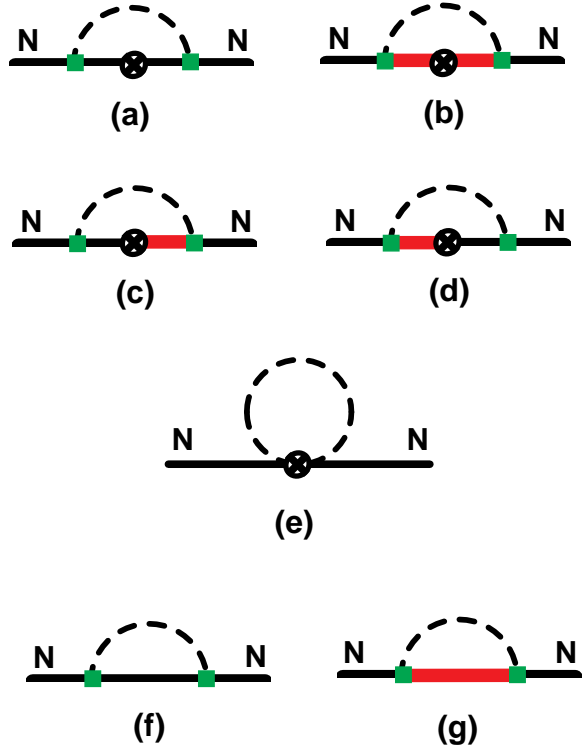}}} 
\vskip 0.15in
\noindent
\caption{\it 
  One-loop graphs that contribute to the matrix elements of the axial
  current in the nucleon.  A solid, thick-solid and dashed line denote
  a nucleon, a $\Delta$-resonance, and a pion, respectively.  The
  solid-squares denote an axial coupling given in
  eq.(\ref{eq:lagstrong}), while the crossed circle denotes an
  insertion of the axial-vector current operator.  Diagrams (a) to (e)
  are vertex corrections, while diagrams (f) and (g) give rise to
  wave-function renormalization.  }
\label{fig:ga}
\vskip .2in
\end{figure}

The nucleon matrix element of the axial current has been extensively
studied in the $p$-expansion, and is known to be~\cite{Jenkins:1991es}
\begin{eqnarray}
\label{eq:gainfinite}
\Gamma_{NN}(\infty) &=& g_A - i {4\over 3 f^2}
\left[ 4 g_A^3\ J_\pi(0)\ +\ 4\left( g_{\Delta N}^2 g_A + {25\over 81}g_{\Delta N}^2
  g_{\Delta\Delta}\right)\ J_\pi(\Delta) \
\right. \\&&\left.\qquad\qquad\qquad
+\ {3\over 2} g_A\ R_\pi \  -\  {32\over 9} g_{\Delta N}^2 g_A\
N_\pi(\Delta)\ \right]
+{\rm counterterms,}
\nonumber
\end{eqnarray}
where the loop contributions, defined with dimensional regularization,
are
\begin{eqnarray}
J_\pi(\Delta) & = & \mu^{4-n}\int {d^nq\over (2\pi)^n}\ 
{(S\cdot q)^2\over (v\cdot q - \Delta  +  i \eta)^2}\ 
{1\over q^2 - m_\pi^2 + i \eta}\,,
\nonumber\\
N_\pi(\Delta) & = & \mu^{4-n}\int {d^nq\over (2\pi)^n}\ 
{(S\cdot q)^2\over v\cdot q - \Delta  +  i \eta}\ 
{1\over v\cdot q   +  i \eta}\ 
{1\over q^2 - m_\pi^2 + i \eta}\,,
\ \ .
\end{eqnarray}
and $R_\pi$ is defined in eq.~(\ref{eq:R1}). These result from the
diagrams shown in Fig.~\ref{fig:ga}, and to be clear, we emphasize
that $g_A$, $g_{N\Delta}$ and $g_{\Delta\Delta}$ are the infinite
volume, chiral limit axial couplings.

Recently, the finite volume corrections in the $m_\pi L\gg 1$ limit of
the $p$-expansion have been computed~\cite{Beane:2004rf}, and are
\begin{eqnarray}
\Gamma_{NN}(L) & = &  \Gamma_{NN}(\infty)
\ +\ 
{m_\pi^2\over 3\pi^2 f^2}\left[\ 
g_A^3 {\bf F_1}
+
\left( g_{\Delta N}^2 g_A + {25\over 81}g_{\Delta N}^2
  g_{\Delta\Delta}\right)  {\bf F_2}
+
g_A  {\bf F_3}
+
g_{\Delta N}^2 g_A {\bf F_4}
\ \right], \nonumber \\
\label{eq:gafinitesize}
\end{eqnarray}
where
\begin{eqnarray}
{\bf F_1} & = & 
\sum_{{\bf n}\ne {\bf 0}}
\left[\ 
K_0({m_\pi} L |{\bf n}|) - {K_1({m_\pi} L |{\bf n}|)\over {m_\pi} L |{\bf n}|}
\ \right]\,,
\nonumber\\
{\bf F_2} & = & 
-\sum_{{\bf n}\ne {\bf 0}}
\left[\ 
{K_1({m_\pi} L |{\bf n}|)\over {m_\pi} L |{\bf n}|}
\ +\ {\Delta^2-{m_\pi}^2\over {m_\pi}^2} K_0({m_\pi} L |{\bf n}|)
\right.\nonumber\\ &&\left.
\qquad -\ 
{\Delta\over {m_\pi}^2}
\int_{m_\pi}^\infty d\beta \ 
{ 2\beta\  K_0(\beta L |{\bf n}|) +(\Delta^2-{m_\pi}^2) L |{\bf n}|\
  K_1(\beta L |{\bf n}|) 
\over \sqrt{\beta^2+\Delta^2-{m_\pi}^2}}
\right]\,,
\label{eq:fsdefined}
\\
{\bf F_3} & = & 
-{3\over 2} \sum_{{\bf n}\ne {\bf 0}} {K_1({m_\pi} L |{\bf n}|)\over {m_\pi} L |{\bf n}|}\,,
\nonumber\\
{\bf F_4} & = & 
{8\over 9} \sum_{{\bf n}\ne {\bf 0}} 
\left[\ 
{K_1({m_\pi} L |{\bf n}|)\over {m_\pi} L |{\bf n}|}
 - 
{\pi e^{-{m_\pi} L |{\bf n}|}\over 2\Delta L |{\bf n}|}
 -
{\Delta^2-{m_\pi}^2\over {m_\pi}^2\Delta}
\int_{m_\pi}^\infty d\beta \ 
{ \beta\  K_0(\beta L |{\bf n}|)\over \sqrt{\beta^2+\Delta^2-{m_\pi}^2}}
\right]\,. \nonumber
\end{eqnarray}
These finite-volume corrections enter at order $p^2$ in the
$p$-expansion, and arise from the one-loop diagrams shown in
Fig.~\ref{fig:ga}.

\subsubsection{ The Axial Current Matrix Element in the Nucleon  in
  the $\epp$-expansion}

The finite-volume contributions to the axial matrix element in the
$\epp$-regime behave differently again from those of both the nucleon
mass and the magnetic moment.  The one-loop diagrams with nucleon or
$\Delta$ intermediate states, diagrams (a), (b), (c), (d), (f) and (g)
of Fig.~\ref{fig:ga}, contribute at order ${\epp}^2$, as the momentum
zero-modes do not contribute due to the derivative coupling to the
baryons. That is,
\begin{eqnarray}
{\rm loop}\sim{1\over L^3}\sum_{{\bf q}}
\int dq_0\ 
{1\over q_0^2 - |{\bf q}|^2}\ \left[ {1\over q_0}\right]^2 \ |{\bf q}|^2
\sim {\epp}^3\ \epp\  {\epp}^{-2}\  {\epp}^{-2}\  {\epp}^2
\ \sim \ {\epp}^2
\ \ ,
\end{eqnarray}
where $q_0\sim |{\bf q}|\sim {\epp}$.  

Pion zero-modes do contribute to the one-loop diagram (diagram (e) of
Fig.~\ref{fig:ga}) resulting from the two pion term in the expansion
of the axial-current insertion (there is no derivative to eliminate
such a contribution).  As for the magnetic moment, this zero-mode
contribution is order ${\epp}$, while the non-zero-mode contributions
remain at order ${\epp}^2$.  Thus, one can determine the leading
volume dependence of the axial-current matrix element directly from
the zero-mode contribution to the function ${\bf F_3}$ in
eq.~(\ref{eq:fsdefined}), and finds that
\begin{eqnarray}
\label{eq:gacorr}
\Gamma_{NN} (L) & = & g_A  \left[\ 1 \ -\ {1\over m_\pi f^2 L^3}\
\right]\ +\  {\cal O}({\epp}^2)
\ \ ,
\end{eqnarray}
in the $\epp$-regime. Again, one cannot take the pion mass to zero at
fixed $L$ in this expression as one rapidly leaves the realm of
applicability of the $\epp$-expansion.

\begin{figure}[!t]
  \centerline{{\epsfxsize=1.5in \epsfbox{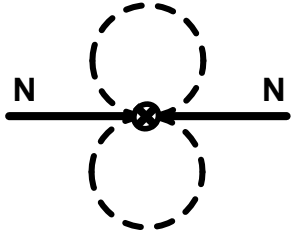}}} \vskip 0.15in
\noindent
\caption{\it 
  A two-loop contribution to $g_A$ that contributes at order
  ${\epp}^2$. This diagram enters at order $p^4$ in the $p$-expansion. }
\label{fig:gatwoloop}
\vskip .2in
\end{figure}

It is tempting to simply take finite-volume loop corrections computed
in the $p$-regime (e.g. eq.~(\ref{eq:gafinitesize})) and look at the
$m_\pi L\alt 1$ limit in an attempt to recover the result in the
$\epp$-regime. However, both the magnetic moment and the axial-current
matrix element demonstrate why this will give incorrect results.  As
discussed above, at order ${\epp}^2$ there will be contributions from
the various one-loop diagrams in Fig.~\ref{fig:ga} and from the
nonzero modes of the two-pion operator tadpole.  In addition, there
will be contributions from the two-loop diagram involving momentum
zero-mode pions that arise from the four pion piece of the
axial-current, Fig.~\ref{fig:gatwoloop}.  Such contributions only
enter at order $p^4$ in the $p$-expansion and are not present in
eq.~(\ref{eq:gafinitesize}).
%

\section{Discussion and Conclusions}

We have defined a power-counting ($\epp$-power-counting) to describe
observables calculated in lattice QCD on highly asymmetric lattices
which are long in the time-dimension and short in the
spatial-dimensions, compared to the inverse pion mass.  In such
volumes, the relative size of contributions from counterterms and loop
diagrams is modified from that of both the $p$-power-counting and the
$\ep$-power-counting.  Loop diagrams involving momentum zero-modes of
the pion field are promoted over those involving the non-zero modes
but still remain perturbative (as opposed to the more general
$\delta$-regime of Leutwyler), while the quark-mass dependent
counterterms are demoted.

One issue that is somewhat uncertain is the range of applicability of
the $\epp$-power-counting.  Looking more closely at the expansion at
the level of Feynman diagrams, it is clear that in the $L_4\to\infty$
limit it is an expansion in factors of
\begin{eqnarray}
\epp & \sim & {m_\pi L\over 2\pi}\ ,\ {\Delta L\over 2\pi}
\ ,\ {2\pi\over \Lambda_\chi L}
\ \ ,
\end{eqnarray}
and the factors of $2\pi$ are important. Of course, these same factors
of $2\pi$ arise in the $\ep$-expansion.  Therefore, the
$\epp$-power-counting applies in the region where $m_\pi L \ll 2\pi$
and $\Lambda_\chi L\gg 2\pi$.  For the physical values of the $m_\pi$
and $\Lambda_\chi$, a lattice dimension of $L=2.5$ -- 4 fm is in the
$\epp$-regime ($\epp<\frac{1}{2}$ in this range). Whilst for
$m_\pi=300$~MeV, boxes $L\sim2.5$~fm are just inside the
$\epp$-regime.

As in the $\ep$-regime, with such narrow windows of applicability and
with a marginally small expansion parameter, it will be important to
extend our analysis to higher orders. Since zero-modes and non-zero
modes are treated on different footings, this rapidly becomes
complicated. A tractable way to formulate the problem is to define
separate fields for the different modes and construct the low energy
effective theory directly in terms of these degrees of freedom. Such
an approach would be similar in spirit to that of soft-collinear
effective theory (e.g., \cite{Bauer:2000yr}), but we do not pursue it
here.

Throughout this work, we have taken $\alpha=\infty$ so that the time
direction is infinite and we have effectively considered a
(Minkowski-space) Hamiltonian formulation of the problem. This was a
convenient simplification and not strictly necessary for our analysis;
all that is required is that $\alpha\gg2$. Provided that this is the
case, corrections to the $L_4\to\infty$ limit will appear perturbatively
in the $\epp$-expansion. Concomitant with the above constraints on $L$
is that $m_\pi L_4\gg2\pi$, so the time direction of $\epp$ lattices is
large, $L_4/L\sim {\epp}^{(1-\alpha)}$.

It is obvious from the above discussion that most (if not all) current
lattice simulations do not fall into the $\epp$-regime. However, since
large finite volume effects have been observed in lattice calculations
of various hadronic quantities including $g_A$
\cite{Sasaki:2003jh,Dolgov:2002zm,Capitani:1999zd}, it is interesting
to insert some numbers.  Taking $m_\pi=300$~MeV and $L=2.5$~fm (which
would require $L_4\agt 6$~fm to be in the $\epp$-regime), the leading
$\epp$ correction to $g_A$ is found to be 10\% from
eq.~(\ref{eq:gacorr}).  That is, a measurement of the axial matrix
element of this volume will give a result 10\% lower than the infinite
volume value at this pion mass.  For the same parameters, the
modification of the nucleon mass is $\sim2$\%. If such large
corrections are present in lattice calculations of $g_A$, they would
go a long way towards explaining the observed discrepancy.

Here we have worked in two-flavour QCD, but it is also interesting to
consider the $\epp$-regime in quenched QCD.  It has long been known
that the singlet propagator of quenched theories leads to strongly
enhanced finite size effects and the same is true in the
$\epp$-expansion. Loops involving the zero-mode of a $\eta^\prime$
meson will contribute at order ${\epp}^0$ and must be treated
non-perturbatively by integrating over certain directions in the coset
space of the graded group \cite{Damgaard:1998xy}.  These modes are
precociously in the $\ep$-regime.

It is interesting to consider two nucleons in the $\epp$-regime. A
fundamental issue in nuclear physics is to understand the quark mass
dependence of nuclear properties and processes. Recent progress
\cite{Beane:2002vq,Beane:2002xf,Epelbaum:2002gb,Epelbaum:2002gk} in
this area has highlighted the fact that quark mass dependent four
nucleon operators in nuclear effective field theories are extremely
difficult to isolate experimentally and and in all likelyhood will
only be determined from lattice QCD. For large volumes where $r\ll L$
(where $r\sim m_\pi^{-1}$ is the range of the nuclear force),
L\"uscher's analysis \cite{Luscher:1986pf} of two particle energy
levels is applicable and one can determine the two-nucleon elastic
scattering parameters \cite{Beane:2003da}. However, to determine the
contributions of the quark mass dependent operators, lattice
calcualtions must be performed over a range of quark masses.
Additionally, quark mass dependent contributions are only suppressed
by one or two orders in the $p$-counting that is valid in this regime
and make an important contribution to the scattering parameters.  In
contrast, in the regime where $\epp$-counting is valid, the
fine-tuning of the two-nucleon sector that leads to the unnaturally
large scattering lengths and the anomalously weak binding of the
deuteron found in nature will not persist. Therefore in this regime
the power-counting of operators in the effective field theory will be
according to their engineering dimensions. Thus the leading mass
dependent operators will be relatively supressed by ${\epp}^4$ and
analysis of two-nucleon energy levels will cleanly determine the quark
mass independent contributions. A potentially serious impediment to
extracting fundamental information about the two-nucleon sector from
this regime is that the ultraviolet cutoff of the two-nucleon
effective theory with dynamical pions is $\Lambda_{NN}\sim300$~MeV
which makes it unlikely that the required hierarchy of length scales
can be established.

Finally, the utility of the $\epp$-regime is not restricted to heavy
objects such as nucleons.  The very large time-dimension allows for
$q_0\sim |{\bf q}|^2$ kinematics to have heavy objects near their
mass-shell, but does not preclude having $q_0\sim |{\bf q}|$, as is
required to have light-particles near mass-shell.  Therefore, in the
analysis of, for instance, the matrix element of the twist-2 isovector
operators in the pion, there is a contribution from the diagram with
four-pions emerging from the operator insertion which contributes at
the one-loop level.  The momentum zero-mode contribution from such a
diagram will be enhanced in the $\epp$-regime over the non-zero-mode
contributions, and it is clear that the leading volume dependence will
be of the same form as that of the axial-current matrix element in the
nucleon, $\sim 1/(m_\pi f^2 L^3)$.

\bigskip\bigskip
\acknowledgments

We would like to thank Silas Beane, Paulo Bedaque, Jiunn-Wei Chen,
Harald Grie\ss{}hammer, David Lin, Gautam Rupak and Steve Sharpe for
useful discussions.  This work is supported in part by the U.S.~Dept.
of Energy under Grant No.~DE-FG03-97ER4014.

\end{document}